\documentclass[11pt,a4paper]{article}
\usepackage{times}
\usepackage{amsfonts,amsmath,amssymb,amsthm}
\usepackage[numbers,sort&compress]{natbib}

\def\fracskip{\mskip 1mu \relax}
\def\nfrac#1#2{{\fracskip#1\fracskip\over\fracskip#2\fracskip}}

\let\frac=\nfrac
\def\pd#1#2{\frac{\partial#1}{\partial#2}}
\def\pdd#1#2#3{\ifx#2#3\pd{^2#1}{#2^2}\else\pd{^2#1}{#2\partial#3}\fi }

\let\Bl=\Bigl \let\Br=\Bigr
\let\BL=\biggl \let\BR=\biggr
\let\bl=\bigl \let\br=\bigr
\def\arb{is an arbitrary constant}
\def\arbs{are arbitrary constants}

\def\arbfs{are arbitrary functions}
\def\Equation#1. {\medbreak{\bfseries\itshape{Equation\kern.3333em\relax#1.}}\enspace\ignorespaces }

\newtheorem{thm}{Theorem}
\newtheoremstyle{remark}{\medskipamount}{\medskipamount}
  {\small\rmfamily}{\parindent}{\footnotesize\sffamily}{.}{.5em}{}
\theoremstyle{remark}
\newtheorem{remark}{Remark}%[section]
\newenvironment{keyword}{\medskip\footnotesize \noindent\textbf{Keywords:} }{\par}

\title{A new method for constructing exact solutions\\ to nonlinear delay partial differential equations}

\author{Andrei D.~Polyanin$^{1}$\thanks{\texttt{polyanin@ipmnet.ru}} \and
Alexei I.~Zhurov$^{1,2}$\thanks{\texttt{zhurovai@cardiff.ac.uk}}\\[6pt]
$^1\,$Institute for Problems in Mechanics, Russian Academy of Sciences,\\
101 Vernadsky Avenue, bldg~1, 119526 Moscow, Russia\\
$^2\,$Cardiff University, Heath Park, Cardiff CF14 4XY, UK}

\begin{document}
\maketitle
\thispagestyle{empty}
\begin{abstract}
We propose a new method for constructing exact solutions to
nonlinear delay reaction--diffusion equations of the form
$$
u_t=ku_{xx}+F(u,w),
$$
where $u=u(x,t)$, $w=u(x,t-\tau)$, and $\tau$ is the delay time. The method
is based on searching for solutions in the form
$u=\sum^N_{n=1}\xi_n(x)\eta_n(t)$, where the functions $\xi_n(x)$ and
$\eta_n(t)$ are determined from additional functional constraints (which are
difference or functional equations) and the original delay partial differential
equation. All of the equations considered contain one or two arbitrary functions
of a single argument. We describe a considerable number of new exact
generalized separable solutions and a few more complex solutions representing
a nonlinear superposition of generalized separable and
traveling wave solutions. All solutions involve free parameters (in some cases,
infinitely many parameters) and so can be suitable for solving certain
problems and testing approximate analytical and numerical methods for nonlinear delay PDEs.
The results are extended to a wide
class of nonlinear partial differential-difference equations involving arbitrary
linear differential operators of any order with respect to the independent
variables $x$ and~$t$ (in particular, this class includes the nonlinear delay
Klein--Gordon equation) as well as to some partial functional differential
equations with time-varying delay.

\begin{keyword}
delay partial differential equations;
delay reaction-diffusion equations;
exact solutions;
generalized separable solutions;
functional differential equations;
time-varying delay;
nonlinear equations;
delay Klein--Gordon equations
\end{keyword}
\end{abstract}

\section{Introduction}

Nonlinear delay partial differential equations and systems of coupled equations arise in
biology, biophysics, biochemistry, chemistry, medicine, control,
climate model theory, ecology, economics, and many other areas (e.g., see the studies
\cite{wu1,trav1,wu2,huang,4,5,trav2,st0,st1,st2,mel} and references in them). It is noteworthy that
similar equations occur in the mathematical theory of artificial neural
networks, whose results are used for signal and image processing as well as in
image recognition problems \cite{st4,st5,st6,st7,st8,st9,st10,st11,7,8}.

The present paper deals with nonlinear delay reaction--diffusion equations
\cite{wu1,wu2,mel,dor1} of the form
\begin{equation}
u_t=ku_{xx}+F(u,w),\qquad w=u(x,t-\tau).
\label{eq:1}
\end{equation}

A number of exact solutions to the heat equation with a nonlinear
source, which is a special case of equation (\ref{eq:1}) without
delay and with $F(u,w)=f(u)$, are listed, for example, in
\cite{dor2,nuc,kud,ibr,pol2,gal,kud2}. Most comprehensive surveys
of exact solutions to this nonlinear equation can be found in the
handbook~\cite{pol1}; it also describes a considerable number
of generalized and functional separable solutions to nonlinear
reaction--diffusion systems of two coupled equations without delay.

The list of known exact solutions to equation (\ref{eq:1}) is quite limited.

In general, equation (\ref{eq:1}) admits traveling-wave solutions, $u=u(\alpha x+\beta t)$. Such solutions are dealt
with in many studies (e.g., see the studies \cite{trav1,wu2,huang,4,5,trav2} and references in them).
%A complete group
%analysis of the nonlinear differential-difference equation (\ref{eq:1}) was carried out in~\cite{mel}. Four equations
%of the form~(\ref{eq:1}), with only one involving an arbitrary function, were found to admit invariant solutions; two
%of these equations had degenerate solutions (linear in~$x$).

A complete group analysis of the non-linear differential-difference
equation~(\ref{eq:1}) was carried out in~\cite{mel}.
Four equations of the form~(\ref{eq:1}) were found to admit invariant
solutions; two of these equations are of limited interest, since
they have degenerate solutions (linear in~$x$). There was only
one equation that involved an arbitrary function
and had a non-degenerate solution:
\begin{equation}
u_t=ku_{xx}+u[-a\ln u+f(wu^{-b})],\quad \ \ b=e^{a\tau},
\label{eq:*}
\end{equation}
where $f(z)$ is an arbitrary function. The exact solution to this
equation found in~\cite{mel} was
\begin{equation}
u=\exp(Cxe^{-at})\varphi(t), \label{eq:**}
\end{equation}
where $C$ is an arbitrary constant and $\varphi(t)$ is a function
satisfying the delay ordinary differential equation
\begin{equation}
\varphi'(t)=\varphi(t)\bigl[C^2ke^{-2at}-a\ln
\varphi(t)+f\bigl(\varphi(t-\tau)\varphi^{-b}(t)\bigr)\bigr].
\label{eq:***}
\end{equation}
The other equation obtained in \cite{mel} that had a
non-degenerate solution coincides, up to notation, with a special
case of equations (\ref{eq:***}), at $f(z)=c_1+c_2\ln z$.

\begin{remark}
It is noteworthy that equation (\ref{eq:*}) is explicitly dependent on the delay
time~$\tau$, which corresponds to a more general kinetic function $F(u,w,\tau)$
than in equation (\ref{eq:1}).
It is only at $a=0$ in (\ref{eq:*}) that we have a kinetic function
explicitly independent of~$\tau$: $F(u,w)=uf(w/u)$; in this case, (\ref{eq:**})
represents a separable solution, $u=e^{Cx}\varphi(t)$.
\end{remark}

In what follows, the term `exact solution' with regard to nonlinear partial
differential-difference equations, including delay partial differential equations,
is used in the following cases:

(i)\enspace the solution is expressible in terms of elementary functions or
in closed form with definite or indefinite integrals;

(ii)\enspace the solution is expressible in terms of solutions to ordinary
differential or ordinary differential-difference equations (or systems of such equations);

(iii)\enspace the solution is expressible in terms of solutions to linear partial
differential equations.

Combinations of cases (i)--(iii) are also allowed.

This definition generalizes the notion of an exact solution used in~\cite{pol1}
with regard to nonlinear partial differential equations.

\begin{remark} %1
Solution methods and various applications of linear and nonlinear ordinary
differential-difference equations, which are much simpler than nonlinear partial
differential-difference equations, can be found, for example,
in~\cite{ofe1,ofe2,ofe3,ofe4,ofe5,ofe6}.
\end{remark}

\section{General description of the functional constraints method}

Consider a wide class of nonlinear delay reaction--diffusion equations:
\begin{equation}
\begin{aligned}
&u_t=ku_{xx}+uf(z)+wg(z)+h(z),\\
&\quad w=u(x,t-\tau),\quad z=z(u,w),
\end{aligned}
\label{eq:1a}
\end{equation}
where $f(z)$, $g(z)$, and $h(z)$ \arbfs\ and $z=z(u,w)$ is a given function. In
addition, we will sometimes consider more complex equations where $f$, $g$,
and $h$ can additionally depend on the independent variables $x$ or/and~$t$
explicitly.

We look for generalized separable solutions of the form
\begin{equation}
u=\sum^N_{n=1}\Phi_n(x)\Psi_n(t),
\label{eq:1b}
\end{equation}
where the functions $\Phi_n(x)$ and $\Psi_n(t)$ are to be determined in the
analysis.

\begin{remark} %2
For nonlinear partial differential equations, various modifications of the method
of generalized separation of variables based on searching for solutions of the
form~(\ref{eq:1b}) are detailed, for example, in~\cite{pol2,gal,pol1}. These
studies also present a large number of equations that admits generalized
separable solutions.

For nonlinear delay partial differential equations of the form (\ref{eq:1a}) that
involve arbitrary functions, the direct application of the method of generalized
separation of variables turns out to be ineffective.
\end{remark}

The new approach pursued in the present paper is based on searching for
generalized separable solutions of the form (\ref{eq:1b}) that satisfy one the
following two additional {\it functional constraints\/}:
\begin{align}
z(u,w)&=p(x),\hphantom{q(t)}\quad w=u(x,t-\tau);\label{eq:2a}\\
z(u,w)&=q(t),\hphantom{p(x)}\quad w=u(x,t-\tau).\label{eq:2b}
\end{align}
These constraints represent difference equations in~$t$ with $x$ playing the
role of a free parameter. The function $z=z(u,w)$ appears in equation
(\ref{eq:1a}) as the argument of the arbitrary functions. The functions $p(x)$
and $q(t)$ are implicitly dependent on $x$ and~$t$ (expressible in terms of
$\Phi_n(x)$ and $\Psi_n(t)$, respectively) and are determined from the
analysis of equations (\ref{eq:2a}) or~(\ref{eq:2b}) taking into
account~(\ref{eq:1b}). It should be emphasized that there is no need to
obtain general solutions of equations (\ref{eq:2a}) or (\ref{eq:2b});
particular solutions will suffice.

A solution to the difference equation (\ref{eq:2a}) or (\ref{eq:2b}), in view
of~(\ref{eq:1b}), determines an allowed form of the exact solution, whose
final representation is subsequently obtained by substituting the resulting
solution into the specific equation~(\ref{eq:1a}).

In what follows, constraints (\ref{eq:2a}) and (\ref{eq:2b}) will be referred to
as a {\it functional constraint of the first kind} and {\it functional constraint of
the second kind}, respectively.

For functional constraint of the first kind (\ref{eq:2a}), the functions
$\Psi_n(t)$ appearing in~(\ref{eq:1b}) are usually chosen in the form
\begin{equation}
\begin{aligned}
\Psi_0(t)&=1,\quad \Psi_1(t)=t,\quad \Psi_n(t)=e^{\lambda_n t},\\
\Psi_n(t)&=e^{\lambda_n t}\cos(\beta_n t),\quad \Psi_n(t)=e^{\lambda_n t}\sin(\beta_n t),
\label{eq:2c}
\end{aligned}
\end{equation}
with the parameters $\lambda_n$ and $\beta_n$ determined from
(\ref{eq:2a}).

\begin{remark} %3
We have introduced the term {\it functional constraint} by analogy with the
term {\it differential constraint}, which is employed in the method of
differential constraints used to seek exact solutions to nonlinear partial
differential equations; for a description of this method and application
examples, see, for example, \cite{pol1,yan,gal2,and}). In more complex
cases, the original equation and functional constraints (\ref{eq:2a}) and
(\ref{eq:2b}) can contain a time-varying delay $\tau=\tau(t)$ instead of the
constant delay~$\tau$; see the comments after
equation~(\ref{eq:325}) in Section~8.
%in this paper!
\end{remark}

Below we give examples of applying the above method to constructing
generalized separable solutions to some equations of the form (\ref{eq:1a}) as
well as more complex nonlinear partial differential equations with delay.

\section{The equation contains one arbitrary function dependent on $w/u$}

\Equation 1. Consider the equation
\begin{equation}
u_t=ku_{xx}+uf(w/u),
\label{eq:2}
\end{equation}
which is a special case of equation (\ref{eq:1a}) with $g=h=0$ and $z=w/u$.

1.1.\enspace The functional constraint of the second kind (\ref{eq:2b}) becomes
\begin{equation}
w/u=q(t),\qquad\quad w=u(x,t-\tau).
\label{eq:3}
\end{equation}

It is clear that the difference equation (\ref{eq:3}) can be satisfied with a
simple separable solution
\begin{equation}
u=\varphi(x)\psi(t),
\label{eq:4}
\end{equation}
which gives $q(t)=\psi(t-\tau)/\psi(t)$. Substituting (\ref{eq:4}) into~(\ref{eq:2})
and separating the variable, we arrive at the following equations for
$\varphi(x)$ and~$\psi(t)$:
\begin{align}
\varphi''&=a\varphi,\label{eq:5}\\
\psi'(t)&=ak\psi(t)+\psi(t)f(\psi(t-\tau)/\psi(t)),\label{eq:6}
\end{align}
where $a$ \arb.

The general solution of equation (\ref{eq:5}) is expressed as
\begin{equation}
\varphi(x)=\begin{cases}
    C_1\cos\bl(\sqrt{|a|}\, x\br)+C_2\sin\bl(\sqrt{|a|}\, x\br)& \text{if \ $a<0$};\\
    C_1\exp\bl(-\sqrt{a}\, x\br)+C_2\exp\bl(\sqrt{a}\, x\br)& \text{if \ $a>0$};\\
    C_1x+C_2& \text{if \ $a=0$,}
    \end{cases}
\label{eq:7}
\end{equation}
where $C_1$ and $C_2$ \arbs.

\begin{remark} %4
Simple separable solutions of the form (\ref{eq:4}) are also admitted by the more general equation
$$
u_t=ku_{xx}+u f(t,w/u),
$$
in which the kinetic function is additionally dependent on~$t$.
\end{remark}

1.2.\enspace In the simple case $p(x)=p_0=\text{const}$, the functional
constraint of the first kind (\ref{eq:2a}) for equation (\ref{eq:2}) is written as
\begin{equation}
w/u=p_0,\qquad w=u(x,t-\tau).
\label{eq:7a}
\end{equation}
Solutions of the difference equation (\ref{eq:7a}) with $p_0>0$ are sought in
the form
\begin{equation}
u=e^{ct}v(x,t),\qquad v(x,t)=v(x,t-\tau),
\label{eq:7b}
\end{equation}
where $c$ \arb\ and $v(x,t)$ is a $\tau$-periodic function. In this case,
$w/u=p_0=e^{-c\tau}$.

Substituting (\ref{eq:7b}) into equation (\ref{eq:2}) yields a linear problem
for determining~$v$:
\begin{equation}
v_t=kv_{xx}+bv,\qquad v(x,t)=v(x,t-\tau),
\label{eq:8}
\end{equation}
where $b=f(e^{-c\tau})-c$.

The general solution to problem (\ref{eq:8}), which will be denoted
$v=V_1(x,t;b)$ for convenience, is expressed as
\begin{align}
&V_1(x,t;b)=\sum^\infty_{n=0}\exp(-\lambda_n x)\bl[A_n\cos(\beta_nt-\gamma_nx)+B_n\sin(\beta_nt-\gamma_nx)\br]\notag\\
&\hphantom{V_1(x,t;b)}\, +\sum^\infty_{n=1}\exp(\lambda_n x)\bl[C_n\cos(\beta_nt+\gamma_nx)+D_n\sin(\beta_nt+\gamma_nx)\br],\label{eq:10}\\
&\beta_n=\frac{2\pi n}\tau,\quad
\lambda_n=\BL(\frac{\sqrt{b^2+\beta_n^2}-b}{2k}\BR)^{\!1/2},\quad
\gamma_n=\BL(\frac{\sqrt{b^2+\beta_n^2}+b}{2k}\BR)^{\!1/2},\label{eq:11}
\end{align}
where $A_n$, $B_n$, $C_n$, and $D_n$ \arbs\ such that the series
(\ref{eq:10}) with (\ref{eq:11}) as well as the derivatives $(V_1)_t$ and
$(V_1)_{xx}$ are convergent; for example, the convergence can be ensured if
one sets $A_n=B_n=C_n=D_n=0$ for $n>N$, where $N$ is some arbitrary
positive integer.

The following special cases can be distinguished:

(i)\enspace $\tau$-periodic (in~$t$) solutions to problem (\ref{eq:8}) decaying as $x\to\infty$
are given by formulas (\ref{eq:10}) and~(\ref{eq:11}) with $A_0=B_0=0$, $C_n=D_n=0$, and $n=1,\,2,\,\dots\,$;

(ii) $\tau$-periodic solutions to problem (\ref{eq:8}) bounded as $x\to\infty$
are given by formulas (\ref{eq:10}) and (\ref{eq:11}) with $C_n=D_n=0$ and $n=1,\,2,\,\dots\,$;

(iii) a stationary solution is given by formulas (\ref{eq:10}) and (\ref{eq:11}) with $A_n=B_n=C_n=D_n=0$ and $n=1,\,2,\,\dots\,\,$.

To sum up, we have an exact solution to equation (\ref{eq:2})
\begin{equation}
u=e^{ct}V_1(x,t;b),\quad b=f(e^{-c\tau})-c,
\label{eq:12}
\end{equation}
where $c$ \arb\ and $V_1(x,t;b)$ is a $\tau$-periodic function determined by formulas (\ref{eq:10}) and (\ref{eq:11}).

1.3.\enspace
Solutions of the difference equation (\ref{eq:7a}) with $p_0<0$ are sought in the form
\begin{equation}
u=e^{ct}v(x,t),\quad v(x,t)=-v(x,t-\tau),
\label{eq:13}
\end{equation}
where $c$ \arb\ and $v(x,t)$ is a $\tau$-antiperiodic function.
In this case, $w/u=p_0=-e^{-c\tau}$.

Substituting (\ref{eq:13}) into (\ref{eq:2}) yields a linear problem for
determining~$v$:
\begin{equation}
v_t=kv_{xx}+bv,\quad v(x,t)=-v(x,t-\tau),
\label{eq:14}
\end{equation}
where $b=f(-e^{-c\tau})-c$.

The general solution to problem (\ref{eq:14}), which will be denoted
$v=V_2(x,t;b)$ for convenience, is expressed as
\begin{align}
&V_2(x,t;b)=\sum^\infty_{n=1}\exp(-\lambda_n x)\bl[A_n\cos(\beta_nt-\gamma_nx)+B_n\sin(\beta_nt-\gamma_nx)\br]\notag\\
&\hphantom{V_2(x,t;b)}\,+\sum^\infty_{n=1}\exp(\lambda_n x)\bl[C_n\cos(\beta_nt+\gamma_nx)+D_n\sin(\beta_nt+\gamma_nx)\br],\label{eq:15}\\
&\beta_n=\frac{\pi(2n-1)}\tau,\quad
\lambda_n=\BL(\frac{\sqrt{b^2+\beta_n^2}-b}{2k}\BR)^{\!1/2},\quad
\gamma_n=\BL(\frac{\sqrt{b^2+\beta_n^2}+b}{2k}\BR)^{\!1/2},\label{eq:16}
\end{align}
where $A_n$, $B_n$, $C_n$, and $D_n$ \arbs\ such that the series
(\ref{eq:15}) with (\ref{eq:16}) as well as the derivatives
$(V_2)_t$ and $(V_2)_{xx}$ are convergent. Solutions to problem
(\ref{eq:14}) decaying as $x\to\infty$ and $\tau$-antiperiodic
in~$t$ are given by formulas (\ref{eq:15}) and (\ref{eq:16}) with
$C_n=D_n=0$ and $n=1,\,2,\,\dots\,\,$.

To sum up, we have an exact solution to equation (\ref{eq:2}):
\begin{equation}
u=e^{ct}V_2(x,t;b),\quad \ b=f(-e^{-c\tau})-c,
\label{eq:17}
\end{equation}
where $c$ \arb\ and $V_2(x,t;b)$ is a $\tau$-antiperiodic
function determined by formulas (\ref{eq:15}) and (\ref{eq:16}).

\begin{remark} %5
Solutions (\ref{eq:10})--(\ref{eq:11}) and (\ref{eq:15})--(\ref{eq:16}) look very similar. However,
the summation in the former solution starts with $n=0$ rather than $n=1$ and the expressions of~$\beta_n$ are different.
\end{remark}

\section{Equations with one arbitrary function dependent on a linear combination of $u$ and $w$}

\Equation 2.
Consider the equation
\begin{equation}
u_t=ku_{xx}+bu+f(u-w),
\label{eq:18}
\end{equation}
which is a special case of equation (\ref{eq:1a}) with $f(z)=b$, $g=0$, and
$z=u-w$ (the function~$h$ has been renamed~$f$).

2.1.\enspace In this case, the functional constraint of the second kind (\ref{eq:2b}) has the form
\begin{equation}
u-w=q(t),\qquad w=u(x,t-\tau).
\label{eq:19}
\end{equation}

It is clear that the additive separable solution
\begin{equation}
u=\varphi(x)+\psi(t)
\label{eq:20}
\end{equation}
satisfies the difference equation (\ref{eq:19}). We have
$q(t)=\psi(t)-\psi(t-\tau)$. Substituting (\ref{eq:20}) into (\ref{eq:18}) and
separating the variables, one arrives at equations for determining $\varphi(x)$
and~$\psi(t)$:
\begin{align}
&k\varphi''_{xx}+b\varphi=a,\label{eq:21}\\
&\psi'_t(t)=b\psi(t)+a+f(\psi(t)-\psi(t-\tau)),
\label{eq:22}
\end{align}
where $a$ \arb.

The general solution to equation (\ref{eq:21}) with $b\not=0$ and $a=0$ is expressed as
\begin{equation}
\varphi(x)=\begin{cases}
    C_1\cos(\alpha x)+C_2\sin(\alpha x), \qquad \ \alpha=\sqrt{b/k}&\ \text{if \ $b>0$};\\
    C_1\exp(-\alpha x)+C_2\exp(\alpha x),\quad \alpha=\sqrt{-b/k}&\ \text{if \ $b<0$},
    \end{cases}
\label{eq:23}
\end{equation}
where $C_1$ and $C_2$ \arbs.
Solution (\ref{eq:20}) with $b>0$ is periodic in the space coordinate~$x$.

The general solution to equation (\ref{eq:21}) with $b=0$ and $a\not=0$ is
\begin{equation}
\varphi(x)=\frac a{2k}\,x^2+C_1x+C_2.
\label{eq:24}
\end{equation}

\begin{remark} %6
Solutions of the form(\ref{eq:20}) are also admitted by the more general equation
$$
u_t=ku_{xx}+b u+f(t,u-w),
$$
in which the kinetic function is additionally dependent on~$t$.
\end{remark}

2.2.\enspace The functional constraint of the first kind (\ref{eq:2a}) for
equation (\ref{eq:18}) has the form
\begin{equation}
u-w=p(x),\qquad\quad w=u(x,t-\tau).
\label{eq:25}
\end{equation}

The difference equation (\ref{eq:19}) can be satisfied, for example, by
choosing the generalized separable solution
\begin{equation}
u=t\varphi(x)+\psi(x),
\label{eq:26}
\end{equation}
which results in $p(x)=\tau\varphi(x)$.

Substituting (\ref{eq:26}) into (\ref{eq:18}) yields ordinary differential equations
for $\varphi(x)$ and~$\psi(x)$:
\begin{align}
&k\varphi''_{xx}+b\varphi=0,\label{eq:27}\\
&k\psi''_{xx}+b\psi +f(\tau\varphi)-\varphi=0.
\label{eq:28}
\end{align}
Equation (\ref{eq:27}) coincides with (\ref{eq:21}) at $a=0$; its solution is
given by formulas~(\ref{eq:23}). The nonhomogeneous constant-coefficient linear
ordinary differential equation (\ref{eq:28}) is easy to integrate.

Using the above solutions (\ref{eq:20}) and (\ref{eq:26}) as well as the theorem below, one can obtain
more complex exact solutions to equation (\ref{eq:18}); these solutions can have any number
of arbitrary parameters.

\begin{thm}[nonlinear superposition of solutions]\label{thm:1}
Suppose $u_0(x,t)$ is a solution to the nonlinear equation (\ref{eq:18}) and
$v=V_1(x,t;b)$ is any $\tau$-periodic solution to the linear heat equation with
source (\ref{eq:8}). Then the sum
\begin{equation}
u=u_0(x,t)+V_1(x,t;b)
\label{eq:29}
\end{equation}
is also a solution to equation (\ref{eq:18}). The general form of the function
$V_1(x,t;b)$ is given by formulas (\ref{eq:10}) and~(\ref{eq:11}).
\end{thm}

\begin{remark} %7
For the nonlinear equation (\ref{eq:18}), the particular solution $u_0(x,t)$ in
formula~(\ref{eq:29}) can be taken in the traveling-wave form
$u_0=u_0(\alpha x+\beta t)$.
\end{remark}

\begin{remark} %8
Formula (\ref{eq:29}) also remains valid for the more general equation
$$
u_t=ku_{xx}+bu+f(x,t,u-w),
$$
in which the kinetic function depends on three arguments. If $f=f(x,u-w)$, then
a stationary solution $u_0=u_0(x)$ can be used as the first term
in~(\ref{eq:29}), while the second term can be taken in the form
(\ref{eq:10})--(\ref{eq:11}). If $f=f(t,u-w)$, then a spatially homogeneous
solution $u_0=u_0(t)$ can be taken for the first term in~(\ref{eq:29}).
\end{remark}

\Equation 3.
Consider the equation
\begin{equation}
u_t=ku_{xx}+bu+f(u-aw),\qquad a>0,
\label{eq:30}
\end{equation}
which is a special case of equation (\ref{eq:1a}) with $f(z)=b$, $g=0$, and
$z=u-aw$; for convenience, the function~$h$ has been renamed~$f$.

3.1.\enspace The functional constraint of the first kind (\ref{eq:2a}) for
equation (\ref{eq:18}) has the form
\begin{equation}
u-aw=p(x),\qquad w=u(x,t-\tau).
\label{eq:31a}
\end{equation}

The difference equation (\ref{eq:31a}) can be satisfied, for example, with the
generalized separable solution
\begin{equation}
u=e^{ct}\varphi(x)+\psi(x),\quad \ c=\frac 1\tau\ln a,
\label{eq:31b}
\end{equation}
which gives $p(x)=(1-a)\psi(x)$.

Substituting (\ref{eq:31b}) into (\ref{eq:30}) leads to ordinary differential
equations for $\varphi(x)$ and~$\psi(x)$:
\begin{align}
k\varphi''+(b-c)\varphi&=0,\label{eq:31c}\\
k\psi''+b\psi+f(\eta)&=0,\quad \eta=(1-a)\psi.\label{eq:31d}
\end{align}

Up to obvious renaming of variables, equation (\ref{eq:31c}) coincides with
(\ref{eq:21}) at \text{$a=0$}; its solution is given by formulas~(\ref{eq:23}) where
$b$ must be substituted by $b-c$.

3.2.\enspace Using the above solution (\ref{eq:31b})--(\ref{eq:31d}) as well
as the theorem below, one can obtain more complex exact solutions to
equation (\ref{eq:30}); these solutions can have any
number of arbitrary free parameters.

\begin{thm}[generalization of Theorem~\ref{thm:1}]
Suppose $u_0(x,t)$ is a solution to the nonlinear equation (\ref{eq:30}) and
$v=V_1(x,t;b)$ is any $\tau$-periodic solution to the linear heat equation with
source (\ref{eq:8}). Then the sum
\begin{equation}
u=u_0(x,t)+e^{ct}V_1(x,t;b-c),\quad c=\frac 1\tau\ln a,
\label{eq:36}
\end{equation}
is also a solution to equation (\ref{eq:30}). The general form of the function
$V_1(x,t;b)$ is given by formulas (\ref{eq:10}) and~(\ref{eq:11}).
\end{thm}

Formula (\ref{eq:36}) makes it possible to obtain a wide class of exact
solutions to the nonlinear equation~(\ref{eq:30}). Apart from (\ref{eq:31b}),
one can take $u_0=u_0(x)$ and $u_0=u_0(t)$ as well as the more general,
traveling wave solution $u_0=\theta(\alpha x+\beta t)$ as the particular solution
$u_0=u_0(x,t)$; the constants $\alpha$ and $\beta$ are arbitrary and the
function $\theta(y)$ satisfies the ordinary differential-difference equation
\begin{equation}
k\alpha^2\theta''(y)-\beta\theta'(y)+b\theta(y)+f\bl(\theta(y)-a\theta(y-\sigma)\br)=0,\quad \ y=\alpha x+\beta t, \ \ \sigma=\beta\tau.
\notag
\end{equation}

\begin{remark} %9
Formula (\ref{eq:36}) also remains valid for the more general equation
$$
u_t=ku_{xx}+bu+f(x,t,u-aw),
$$
in which the kinetic function depends on three arguments.
\end{remark}

\Equation 4.
Consider the equation
\begin{equation}
u_t=ku_{xx}+bu+f(u+aw),\qquad a>0,
\label{eq:40}
\end{equation}
which is a special case of equation (\ref{eq:1a}) with $f(z)=b$, $g=0$, and
$z=u+aw$; for convenience, the function~$h$ has been renamed~$f$.

\begin{thm} %3
Suppose $u_0(x,t)$ is a solution to the nonlinear equation
(\ref{eq:40}) and $v=V_2(x,t;b)$ is any $\tau$-antiperiodic
solution to the linear heat equation with source (\ref{eq:14}).
Then the sum
\begin{equation}
u=u_0(x,t)+e^{ct}V_2(x,t;b-c),\quad c=\frac 1\tau\ln a,
\label{eq:41}
\end{equation}
is also a solution to equation (\ref{eq:40}). The general form of the function
$V_2(x,t;b)$ is given by formulas (\ref{eq:15}) and~(\ref{eq:16}).
\end{thm}

Formula (\ref{eq:41}) makes it possible to obtain a wide class of exact
solutions to the nonlinear equation~(\ref{eq:40}). One can take $u_0=u_0(x)$
and $u_0=u_0(t)$ as well as the more general, traveling wave solution
$u_0=\theta(\alpha x+\beta t)$ as the particular solution $u_0(x,t)$; the
constants $\alpha$ and $\beta$ are arbitrary and the function~$\theta(y)$
satisfies the ordinary differential-difference equation
\begin{equation}
k\alpha^2\theta''(y)-\beta\theta'(y)+b\theta(y)+f\bl(\theta(y)+a\theta(y-\sigma)\br)=0,\quad \ y=\alpha x+\beta t, \ \ \sigma=\beta\tau.
\notag
\end{equation}

\begin{remark} %10
Formula (\ref{eq:41}) also remains valid for the more general equation
$$
u_t=ku_{xx}+bu+f(x,t,u+aw),
$$
in which the kinetic function depends on three arguments.
\end{remark}

\section{Equations with two arbitrary functions dependent on a linear combination of $u$ and $w$}

\Equation 5.
Now consider the more general equation
\begin{equation}
u_t=ku_{xx}+uf(u-w)+wg(u-w)+h(u-w),
\label{eq:32}
\end{equation}
where $f(z)$, $g(z)$, and $h(z)$ \arbfs ; in this case, either function $f$ or~$g$
can be set equal to zero without loss of generality.

5.1.\enspace The difference constraint (\ref{eq:2a}) for equation (\ref{eq:32})
has the form (\ref{eq:25}). The linear difference equation~(\ref{eq:25}) can
be satisfied, as previously, with a generalized separable solution of the
form~(\ref{eq:26}). As a result, one can obtain equations for determining
$\varphi(x)$ and~$\psi(x)$; these equations are not written out, since a
significantly more general result will be presented below.

5.2.\enspace
The linear difference equation (\ref{eq:25}) can be satisfied by setting
\begin{equation}
u=\sum^N_{n=1}[\varphi_n(x)\cos(\beta_nt)+\psi_n(x)\sin(\beta_nt)]+t\theta(x)+\xi(x),\quad \beta_n=\frac{2\pi n}\tau,
\label{eq:33}
\end{equation}
where $N$ is an arbitrary positive integer. In this case, we have
$p(x)=\tau\varphi(x)$ on the right-hand side of equation~(\ref{eq:25}).

Substituting (\ref{eq:33}) into (\ref{eq:32}) and performing simple rearrangements,
we obtain
\begin{equation}
\sum^N_{n=1}[A_n\cos(\beta_nt)+B_n\sin(\beta_nt)]+Ct+D=0,
\label{eq:34}
\end{equation}
where the functional coefficients $A_n$,~$B_n$,~$C$, and~$D$ are
dependent on $\varphi_n(x)$, $\psi_n(x)$, $\theta(x)$, and~$\xi(x)$ as well
as their derivatives and independent of~$t$. In~(\ref{eq:34}), equating all
the functional coefficients with zero, $A_n=B_n=C=D=0$, we arrive at the
following ordinary differential equation for the unknown functions:
\begin{align*}
k\varphi_n''+\varphi_n[f(\tau\theta)+g(\tau\theta)]-\beta_n\psi_n&=0,\\
k\psi_n''+\psi_n[f(\tau\theta)+g(\tau\theta)]+\beta_n\varphi_n&=0,\\
k\theta''+\theta[f(\tau\theta)+g(\tau\theta)]&=0,\\
k\xi''+\xi f(\tau\theta)+(\xi-\tau\theta)g(\tau\theta)+h(\tau\theta)-\theta&=0.
\end{align*}

Note that the third nonlinear equation admits the trivial particular solution
$\theta=0$; in this case, all other equations become linear with constant
coefficients.

\begin{remark} %11
Solutions of the form (\ref{eq:33}) are also admitted by the more general
equation
$$
u_t=ku_{xx}+uf(x,u-w)+wg(x,u-w)+h(x,u-w).
$$
\end{remark}

\Equation 6.
Consider the equation
\begin{equation}
u_t=ku_{xx}+uf(u-aw)+wg(u-aw)+h(u-aw),
\label{eq:34a}
\end{equation}
where $f(z)$, $g(z)$, and $h(z)$ \arbfs, which generalizes
equation~(\ref{eq:30}).

6.1.\enspace
The difference constraint (\ref{eq:2a}) for equation (\ref{eq:34a})
has the form (\ref{eq:31a}). The linear difference equation~(\ref{eq:31a})
can be satisfied, as previously, with a generalized separable solution of the form~(\ref{eq:31b}).
As a result, one arrives at equations for determining $\varphi(x)$ and $\psi(x)$;
these equations are not written here, since a much more general result is given below.

6.2.\enspace
The linear difference equation (\ref{eq:25}) can be satisfied by setting
\begin{align}
u&=e^{ct}\BL\{\theta(x)+\sum^N_{n=1}[\varphi_n(x)\cos(\beta_nt)+\psi_n(x)\sin(\beta_nt)]\BR\}+\xi(x),\label{eq:39}\\
c&=\frac 1\tau\ln a,\quad \beta_n=\frac{2\pi n}\tau,\notag
\end{align}
where $N$ is an arbitrary positive integer. In this case, we have
$p(x)=(1-a)\xi(x)$ of the right-hand side of equation~(\ref{eq:31a}).

Substituting (\ref{eq:39}) into (\ref{eq:34a}) and reasoning in a similar
fashion as for equation~(\ref{eq:32}), we arrive at the following equations for
determining $\theta(x)$, $\varphi_n(x)$, $\psi_n(x)$, and~$\xi(x)$:
\begin{align*}
&k\theta''+\theta\Bl[f(\eta)+\frac 1a g(\eta)-c\Br]=0,\quad \eta=(1-a)\xi,\\
&k\varphi_n''+\varphi_n\Bl[f(\eta)+\frac 1a g(\eta)-c\Br]-\beta_n\psi_n=0,\\
&k\psi_n''+\psi_n\Bl[f(\eta)+\frac 1a g(\eta)-c\Br]+\beta_n\varphi_n=0,\\
&k\xi''+\xi[f(\eta)+g(\eta)]+h(\eta)=0.
\end{align*}

\begin{remark} %12
Solutions of the form (\ref{eq:39}) are also admitted by the more general
equation
$$
u_t=ku_{xx}+uf(x,u-aw)+wg(x,u-aw)+h(x,u-aw),\qquad a>0.
$$
\end{remark}

\Equation 7.
Consider the equation
\begin{equation}
u_t=ku_{xx}+uf(u+aw)+wg(u+aw)+h(u+aw),\qquad a>0,
\label{eq:35}
\end{equation}
where $f(z)$, $g(z)$, and $h(z)$ \arbfs, which generalizes equation
(\ref{eq:40}).

The difference constraint (\ref{eq:2a}) for equation (\ref{eq:35}) has the form
\begin{equation}
u+aw=p(x),\qquad\quad w=u(x,t-\tau).
\label{eq:35a}
\end{equation}

The linear difference equation (\ref{eq:35a}) can be satisfied by setting
\begin{align}
u&=e^{c t}\sum^N_{n=1}[\varphi_n(x)\cos(\beta_nt)+\psi_n(x)\sin(\beta_nt)]+\xi(x),\label{eq:35b}\\
c&=\frac 1\tau\ln a,\quad \beta_n=\frac{\pi(2n+1)}\tau,\notag
\end{align}
where $N$ is an arbitrary positive integer. In this case, we have
$p(x)=(1+a)\xi(x)$ on the right-hand side of equation~(\ref{eq:35a}).

Substituting (\ref{eq:35b}) into (\ref{eq:35}) and reasoning in a similar
fashion as for equation~(\ref{eq:32}), we arrive at the following equations for
determining $\varphi_n(x)$, $\psi_n(x)$, and~$\xi(x)$:
\begin{align*}
&k\varphi_n''+\varphi_n\Bl[f(\eta)-\frac 1a g(\eta)-c\Br]-\beta_n\psi_n=0,\\
&k\psi_n''+\psi_n\Bl[f(\eta)-\frac 1a g(\eta)-c\Br]+\beta_n\varphi_n=0,\\
&k\xi''+\xi[f(\eta)+g(\eta)]+h(\eta)=0,\quad \eta=(1+a)\xi.
\end{align*}
The last equation is independent.

\begin{remark} %13
Solutions of the form (\ref{eq:35b}) are also admitted by the more general equation
$$
u_t=ku_{xx}+uf(x,u+aw)+wg(x,u+aw)+h(x,u+aw),\qquad a>0.
$$
\end{remark}

\section{Equation with two arbitrary functions dependent on $u^2+w^2$}

\Equation 8. Now consider the equation
\begin{equation}
u_t=ku_{xx}+uf(u^2+w^2)+wg(u^2+w^2),
\label{eq:100}
\end{equation}
which contains a nonlinear (quadratic) argument, $z=u^2+w^2$.

The difference constraint (\ref{eq:2a}) for equation (\ref{eq:100})
has the form
\begin{equation}
u^2+w^2=p(x),\qquad w=u(x,t-\tau).
\label{eq:101}
\end{equation}

The nonlinear difference equation (\ref{eq:35a}) can be satisfied by setting
\begin{align}
&u=\varphi_n(x)\cos(\lambda_n t)+\psi_n(x)\sin(\lambda_n t),\label{eq:102}\\
&\lambda_n=\frac{\pi(2n+1)}{2\tau}, \quad n=0,\,\pm1,\,\pm2,\,\dots\,.\notag
\end{align}
It is not difficult to verify that
$$
w=(-1)^n\varphi_n(x)\sin(\lambda_n t)+(-1)^{n+1}\psi_n(x)\cos(\lambda_n t)
$$
and
$$
u^2+w^2=\varphi_n^2(x)+\psi_n^2(x)=p(x).
$$

Substituting (\ref{eq:102}) into (\ref{eq:100}) and splitting the resulting
expression with respect to $\cos(\lambda_n t)$ and $\sin(\lambda_n t)$, we
arrive at a nonlinear system of ordinary differential equations for determining
$\varphi_n(x)$ and~$\psi_n(x)$:
\begin{align}
k\varphi_n''+\varphi_n f(\varphi_n^2+\psi_n^2)+(-1)^{n+1}\psi_n g(\varphi_n^2+\psi_n^2)-\lambda_n\psi_n&=0,\notag\\
k\psi_n''+\psi_n f(\varphi_n^2+\psi_n^2)+(-1)^{n}\varphi_n g(\varphi_n^2+\psi_n^2)+\lambda_n\varphi_n&=0.\notag
\end{align}

\begin{remark} %14
Solutions of the form (\ref{eq:35b}) are also admitted by the more general equation
$$
u_t=ku_{xx}+uf(x,u^2+w^2)+wg(x,u^2+w^2).
$$
\end{remark}

\section{More general nonlinear delay partial differential equations}

Now consider nonlinear partial differential-difference equations of the more general form
\begin{equation}
\begin{aligned}
\text{L}[u]&=\text{M}[u]+uf(z)+wg(z)+h(z),\\
w&=u(x,t-\tau),\quad z=z(u,w),
\end{aligned}
\label{eq:301}
\end{equation}
where
$\text{L}$ and $\text{M}$ are arbitrary constant-coefficient linear differential operators with respect to $t$ and $x$:
\begin{equation}
\text{L}[u]=\sum^k_{i=1}b_i\pd{^iu}{t^i},\quad
\text{M}[u]=\sum^m_{i=1}a_i\pd{^iu}{x^i}.
\label{eq:302}
\end{equation}
As before, $f(z)$, $g(z)$, and $h(z)$ \arbfs\ and $z=z(u,w)$ is a given
function.

By setting $\text{L}[u]=u_{tt}$ and $\text{M}[u]=a^2u_{xx}$ in
(\ref{eq:301}), we get the nonlinear delay Klein--Gordon equation (a delay hyperbolic PDE)
\begin{equation}
u_{tt}=a^2u_{xx}+uf(z)+wg(z)+h(z).
\label{eq:303}
\end{equation}

\begin{remark}
Oscillation properties of some hyperbolic partial differential equations
with delay were studied, for example, in~\cite{hypeqs1,hypeqs2,hypeqs3}.
\end{remark}

Many of the results obtained previously for nonlinear delay reaction--diffusion
equations (\ref{eq:1a}) also apply to nonlinear delay partial differential
equations of the general form (\ref{eq:301})--(\ref{eq:302}).

Subsequently, in deriving determining equations, we use the following simple
properties of the operators $\text{L}$ and~$\text{M}$:
\begin{align*}
\text{L}[\varphi(x)\psi(t)]&=\varphi(x)\text{L}[\psi(t)],\quad \
\text{L}[\varphi(x)+\psi(t)]=\text{L}[\psi(t)],\\
\text{M}[\varphi(x)\psi(t)]&=\psi(t)\text{M}[\varphi(x)],\quad
\text{M}[\varphi(x)+\psi(t)]=\text{M}[\varphi(x)].
\end{align*}
Below we omit intermediate steps and give only final results.

\Equation 9.
The equation
\begin{equation}
\text{L}[u]=\text{M}[u]+u f(w/u),
\label{eq:304}
\end{equation}
admits the multiplicative separable solution
\begin{equation}
u=\varphi(x)\psi(t),
\label{eq:305}
\end{equation}
with the functions $\varphi(x)$ and $\psi(t)$ satisfying the linear ordinary
differential and differential-difference equations
\begin{align}
&\text{M}[\varphi]=c\varphi,\label{eq:306}\\
&\text{L}[\psi(t)]=c\psi(t)+\psi(t)f\bl(t,\psi(t-\tau)/\psi(t)\br),\label{eq:307}
\end{align}
where $c$ \arb.

It should be noted that:

(i)\enspace if $c=0$, equation (\ref{eq:306}) admits polynomial particular solutions;

(ii)\enspace for any $c$, equation (\ref{eq:306}) admits exponential particular
solutions $\varphi(x)=Ae^{\beta x}$, where $A$ \arb\ and $\beta$ is a root of
the polynomial equation $\sum^m_{i=1}a_i\beta^i=c$;

(iii)\enspace equation (\ref{eq:307}) admits exponential particular solutions
$\psi(t)=Be^{\lambda t}$, where $B$ \arb\ and $\lambda$ is a root of the
transcendental equation
$$
\sum^n_{i=1}b_i\lambda^i=c+f(e^{-\tau\lambda}).
$$

\Equation 10.
The equation
\begin{equation}
\text{L}[u]=\text{M}[u]+b u+f(u-w)
\label{eq:309}
\end{equation}
admits additive separable solutions of the form
\begin{equation}
u=\varphi(x)+\psi(t),
\label{eq:310}
\end{equation}
with the functions $\varphi(x)$ and $\psi(t)$ described by the linear
partial differential equation and ordinary differential-difference equation
\begin{align}
&\text{M}[\varphi]=c-b \varphi,\label{eq:311}\\
&\text{L}[\psi(t)]=c+b\psi(t)+f\bl(\psi(t)-\psi(t-\tau)\br)\label{eq:312}
\end{align}
and $c$ being an arbitrary constant.

It should be noted that:

(i)\enspace if $b=0$, equation (\ref{eq:311}) admits polynomial particular solutions;

(ii)\enspace if $c=0$, equation (\ref{eq:311}) admits exponential particular
solutions $\varphi(x)=Ae^{\beta x}$, where $A$ \arb\ and $\beta$ is a root of
the polynomial equation $\sum^n_{i=1}a_i\beta^i=-b$;

Using the above solution (\ref{eq:310}) and the theorem
below, one can find more complex exact solutions to equation~(\ref{eq:309});
these solutions can have any number of arbitrary free parameters.

\begin{thm}[nonlinear superposition of solutions]\label{thm:4}
Suppose $u_0(x,t)$ is a solution to the nonlinear equation~(\ref{eq:309}) and
$v=v(x,t)$ is any $\tau$-periodic solution to the linear equation
\begin{equation}
\text{\rm L}[v]=\text{\rm M}[v]+bv,\quad \ v(x,t)=v(x,t-\tau).
\label{eq:311b}
\end{equation}
Then the sum
\begin{equation}
u=u_0(x,t)+v(x,t)
\label{eq:312b}
\end{equation}
is also solution to equation (\ref{eq:309}).
\end{thm}

Apart from (\ref{eq:310}), the particular solution $u_0(x,t)$ can be sought in
the traveling-wave form $u_0=\theta(\alpha x+\beta t)$. Particular solutions
to the linear problem (\ref{eq:311b}) can be sought in the form
\begin{align}
&v(x,t)=\sum^N_{n=0}\exp(-\lambda_n x)\bl[A_n\cos(\beta_nt-\gamma_nx)+B_n\sin(\beta_nt-\gamma_nx)\br]\notag\\
&\hphantom{v(x,t)}\,+\sum^N_{n=1}\exp(\lambda_n x)\bl[C_n\cos(\beta_nt+\gamma_nx)+D_n\sin(\beta_nt+\gamma_nx)\br],\label{eq:313}\\
&\beta_n={2\pi n}/\tau,\notag
\end{align}
where $N$ is any positive integer, $A_n$, $B_n$, $C_n$, and $D_n$ \arbs, and
$\lambda_n$ and $\gamma_n$ are constants determined from substituting
(\ref{eq:313}) into equation~(\ref{eq:311b}).

\Equation 11.
For the equation
\begin{equation}
\text{L}[u]=\text{M}[u]+bu+f(u-aw),\qquad a>0,
\label{eq:314}
\end{equation}
the following theorem holds.

\begin{thm}[generalizes Theorem \ref{thm:4}]
Suppose $u_0(x,t)$ is a solution to the nonlinear equation~(\ref{eq:314}) and
$v=v(x,t)$ is any $\tau$-periodic solution to the linear equation
\begin{equation}
\text{\rm L}_1[v]=\text{\rm M}[v]+bv,
\quad \ v(x,t)=v(x,t-\tau),
\label{eq:315}
\end{equation}
where $\text{\rm L}_1[v]\equiv e^{-ct}\text{\rm L}[e^{ct}v]$ is a linear differential
operator with constant coefficients. Then the sum
\begin{equation}
u=u_0(x,t)+e^{ct}v(x,t),\quad c=\frac 1\tau\ln a,
\label{eq:316}
\end{equation}
is also a solution to equation (\ref{eq:314}).
\end{thm}

Formula (\ref{eq:316}) makes it possible to obtain a wide class of exact
solutions to the nonlinear equation~(\ref{eq:314}). Simple solutions of the
form $u_0=u_0(x)$ and $u_0=u_0(t)$ as well as the more general, traveling
wave solution $u_0=\theta(\alpha x+\beta t)$ can be taken as the particular
solution $u_0(x,t)$. Just as before, particular solutions to the linear
problem~(\ref{eq:315}) can be sought in the form (\ref{eq:313}).

\Equation 12.
For the equation
\begin{equation}
\text{L}[u]=\text{M}[u]+bu+f(u+aw),\qquad a>0,
\label{eq:317}
\end{equation}
the following theorem holds.

\begin{thm}
Suppose $u_0(x,t)$ is a solution to the nonlinear
equation~(\ref{eq:317}) and $v=v(x,t)$ is any $\tau$-antiperiodic
solution to the linear equation
\begin{equation}
\text{\rm L}_1[v]=\text{\rm M}[v]+bv,\quad \ v(x,t)=-v(x,t-\tau),
\label{eq:318}
\end{equation}
where $\text{\rm L}_1[v]\equiv e^{-ct}\text{\rm L}[e^{ct}v]$ is a linear differential
operator with constant coefficients. Then the sum
\begin{equation}
u=u_0(x,t)+e^{ct}v(x,t),\quad c=\frac 1\tau\ln a,
\label{eq:319}
\end{equation}
is also a solution to equation (\ref{eq:317}).
\end{thm}

Formula (\ref{eq:319}) makes it possible to obtain a wide class of exact
solutions to the nonlinear equation (\ref{eq:317}). Simple solutions of the form
$u_0=u_0(x)$ and $u_0=u_0(t)$ as well as the more general, traveling wave
solution $u_0=\theta(\alpha x+\beta t)$ can be taken as the particular solution
$u_0(x,t)$. Particular solutions to the linear problem~(\ref{eq:318}) can be
sought in the form
\begin{align}
&v(x,t)=\sum^N_{n=1}\exp(-\lambda_n x)\bl[A_n\cos(\beta_nt-\gamma_nx)+B_n\sin(\beta_nt-\gamma_nx)\br]\notag\\
&\hphantom{v(x,t)}\,+\sum^N_{n=1}\exp(\lambda_n x)\bl[C_n\cos(\beta_nt+\gamma_nx)+D_n\sin(\beta_nt+\gamma_nx)\br],\label{eq:320}\\
&\beta_n=\frac{\pi(2n-1)}\tau,\notag
\end{align}
where $N$ is any positive integer, $A_n$, $B_n$, $C_n$, and $D_n$ \arbs, and
$\lambda_n$ and $\gamma_n$ are constants determined from substituting
(\ref{eq:320}) into~(\ref{eq:318}).

\Equation 13.
Exact solutions to the nonlinear delay partial differential equation
\begin{equation}
\text{L}[u]=\text{M}[u]+uf(x,u-w)+wg(x,u-w)+h(x,u-w)
\label{eq:321}
\end{equation}
are sought in the form (\ref{eq:33}).

\Equation 14.
Exact solutions to the nonlinear delay partial differential equation
\begin{equation}
\text{L}[u]=\text{M}[u]+uf(x,u-aw)+wg(x,u-aw)+h(x,u-aw),\quad a>0,
\label{eq:322}
\end{equation}
are sought in the form (\ref{eq:39}).

\Equation 15.
Exact solutions to the nonlinear delay partial differential equation
\begin{equation}
\text{L}[u]=\text{M}[u]+uf(x,u+aw)+wg(x,u+aw)+h(x,u+aw),\quad a>0,
\label{eq:323}
\end{equation}
are sought in the form (\ref{eq:35b}).

\Equation 16.
Exact solutions to the nonlinear delay partial differential equation
\begin{equation}
\text{L}[u]=\text{M}[u]+uf(x,u^2+w^2)+wg(x,u^2+w^2)
\label{eq:324}
\end{equation}
are sought in the form (\ref{eq:102}).

\begin{remark} %15
In equations (\ref{eq:321}), (\ref{eq:322}), (\ref{eq:323}), and (\ref{eq:324}),
the linear operator~$\text{M}$ can depend on~$x$ and have the form
\begin{equation*}
\text{M}[u]=\sum^m_{i=1}a_i(x)\pd{^iu}{x^i},
\end{equation*}
where $a_i(x)$ \arbfs.
\end{remark}

\begin{remark} %16
In equations (\ref{eq:321}), (\ref{eq:322}), (\ref{eq:323}),
and~(\ref{eq:324}), the functions $f$,~$g$, and~$h$ can explicitly depend on
several space coordinates, $x=(x_1,\dots,x_s)$, and $\text{M}$ can be an
arbitrary linear differential operator of any order (first, second, or higher) with
respect to the space coordinates whose coefficients can be explicitly dependent
on $x=(x_1,\dots,x_s)$. Just as before, exact solutions to such equations are
sought in the form (\ref{eq:33}), (\ref{eq:39}), (\ref{eq:35b}), and
(\ref{eq:102}), respectively, where $x$ is understood as $x=(x_1,\dots,x_s)$.

In particular, one can choose an elliptic operator,
\begin{equation*}
\text{M}[u]=\sum^s_{i,j=1}\pd{}{x_i}k_{ij}(\textbf{x})\pd u{x_j}+\sum^s_{i=1}c_{i}(\textbf{x})\pd u{x_i},
\end{equation*}
which arises in the studies \cite{7,8} in the case $c_i=0$.
\end{remark}

\section{General partial differential equations with time-varying delay}

In very much the same manner, one can obtain exact solutions to
some nonlinear partial functional differential equations
with time-varying delay of general form.

Below we give a few simple examples to illustrate the aforesaid.

\Equation 17.
Consider nonlinear partial functional differential equations of the form
\begin{equation}
\text{L}[u]=\text{M}[u]+u f(w/u),\qquad w=u(x,\zeta(t)),
\label{eq:325}
\end{equation}
where $\zeta(t)$ is a given function. In applications (e.g., see~\cite{7}), it is
conventional to write $\zeta(t)=t-\tau(t)$, where $\tau(t)$ is a time-varying
delay such that $0\le\tau(t)<\tau_0$.

In this case, the functional constraint (\ref{eq:2b}) becomes
\begin{equation}
w/u=q(t),\quad w=u(x,\zeta(t)).
\label{eq:326}
\end{equation}

The functional equation (\ref{eq:326}) can be satisfied with
the simple multiplicative separable solution
\begin{equation}
u=\varphi(x)\psi(t),
\label{eq:327}
\end{equation}
which gives $q(t)=\psi(\zeta(t))/\psi(t)$. Substituting (\ref{eq:327}) into
(\ref{eq:325}) and separating the variables, one arrives at the following
equations for $\varphi(x)$ and~$\psi(t)$:
\begin{align*}
&\text{M}[\varphi]=c\varphi,\\
&\text{L}[\psi(t)]=c\psi(t)+\psi(t)f(t,\psi(\zeta)/\psi(t)),\quad \zeta=\zeta(t),
\end{align*}
where $c$ \arb.

\Equation 18.
Consider nonlinear partial functional differential equations of the form
\begin{equation}
\text{L}[u]=\text{M}[u]+b u+f(u-w),\qquad w=u(x,\zeta(t)).
\label{eq:328}
\end{equation}

In this case, the functional constraint (\ref{eq:2b}) becomes
\begin{equation}
u-w=q(t),\quad w=u(x,\zeta(t)).
\label{eq:329}
\end{equation}
The equation admits an additive separable solution of the form
\begin{equation}
u=\varphi(x)+\psi(t),
\label{eq:330}
\end{equation}
which gives $q(t)=\psi(t)-\psi(\zeta(t))$. Substituting (\ref{eq:330}) into
(\ref{eq:328}) and separating the variables, one arrives at the following
equations for $\varphi(x)$ and~$\psi(t)$:
\begin{align*}
&\text{M}[\varphi]=c-b \varphi,\\
&\text{L}[\psi(t)]=c+b\psi(t)+f(\psi(t)-\psi(\zeta)),\quad \zeta=\zeta(t),
\end{align*}
where $c$ \arb.

\section{Brief conclusions}

To summarize, we have proposed a new method, which we call the functional
constraints method, for constructing exact solutions of nonlinear delay
reaction--diffusion equations of the form
$$
u_t=ku_{xx}+F(u,w),
$$
where $u=u(x,t)$, $w=u(x,t-\tau)$, and $\tau$ is the relaxation time. The
method is based on searching for generalized separable solutions of the form
$$
u=\sum^N_{n=1}\xi_n(x)\eta_n(t),
$$
with the functions $\xi_n(x)$ and $\eta_n(t)$ determined from additional
functional constraints (difference or functional equations)
and the original delay partial differential equation.

All of the equations considered contain one or two arbitrary functions
of a single argument. We have described a considerable number of new exact
generalized separable solutions and a few more complex solutions representing
a nonlinear superposition of generalized separable solutions and
traveling wave solutions.

All solutions involve free parameters (in some cases, infinitely many
parameters) and so can be suitable for solving certain problems and
testing approximate analytical and numerical methods for nonlinear delay PDEs.

The results have been extended to the following classes of equations:

(i)\enspace nonlinear delay partial differential equations of the form
$$
\text{L}[u]=\text{M}[u]+F(u,w),
$$
where $\text{L}$ and $\text{M}$ are arbitrary constant-coefficient linear
differential operators of any order with respect to the independent variables
$x$ and~$t$; in particular, this broad class includes the nonlinear delay
Klein--Gordon equation, which is a hyperbolic-type delay equation;

(ii)\enspace some nonlinear delay partial differential equations that are
explicitly dependent on either of the independent variables, $x$~or~$t$; and

(iii)\enspace some nonlinear partial functional differential equations with
time-varying delay of general form.

\end{document}